\documentstyle[preprint,aps,eqsecnum]{revtex}
\begin{document}

\title
{Magnetoresistance of Two-Dimensional Fermions in \\a Random Magnetic Field}

\author{D.V. Khveshchenko}
\address
{NORDITA, Blegdamsvej 17, Copenhagen DK-2100, Denmark}

\maketitle

\begin{abstract}
\noindent

We perform a semiclassical calculation of the magnetoresistance of spinless two-dimensional fermions in a long-range correlated random magnetic field. In the regime relevant for
the problem of the half filled Landau level the perturbative Born approximation fails and we develop a new 
method of solving the Boltzmann equation beyond the relaxation time approximation.  
In absence of interactions, electron density modulations, in-plane fields,
and Fermi surface anisotropy we obtain a quadratic negative magnetoresistance in the weak field limit.

\end{abstract}
\pagebreak

The problem of two-dimensional transport in spatially random static magnetic fields (RMF) attracted a lot
of attention over last few years. There exists a number of experimental realizations of two-dimensional 
(2D) RMF created by randomly pinned flux vortices in a type-II superconducting gate, grains of 
a type-I superconductor, or a demagnetized permanent magnet 
placed in the nearby of the 2D electron gas \cite{exp}.
A formally similar problem arises in the contexts of the gauge theory of high $T_c$ cuprates \cite{LN}
and the composite fermion theory of the half filled Landau level (HFLL) \cite{HLR},\cite{KZ}. 

Various analytical and numerical results obtained so far seem to indicate that in the case of a continuum system with 
unbound spectrum all states are localized \cite{loc} while in the lattice case there might be an extended state
at the center of the band \cite{miller}. The effective low-energy  description in terms of the unitary $\sigma$-model constructed in \cite{AMW} suggests that the RMF problem belongs to the unitary
random ensemble. Compared to the conventional problem of potential scattering corresponding to the orthogonal
case,  
 the effects of quantum interference in the RMF  
are suppressed as a result of broken time-reversal symmetry. In particular,
the logarithmic temperature dependent weak localization corrections to the conductivity 
appear only in the next order in the metallicity parameter $k_F l>>1$: 
$\delta_{wl}\sigma = -{1\over {\pi^2 k_F l}}\log {\tau_{\varphi}\over \tau_{tr}}$ (where the temperature dependence comes from the inelastic phase breaking time $\tau_{\varphi}(T)$ in the regime 
$\tau_{\varphi}(T)>>\tau_{tr}={l\over v_F}$). 
It suggests a larger localization length
$L_{loc}\sim l \exp({\pi^2\over 4} (k_F l)^2)$ than in the orthogonal case. 

It was shown in \cite{IA} that quantum 
fluctuations of particle's positions in the direction transverse to their classical trajectories are strongly reduced
by the RMF.
And so is the contribution of quantum backscattering which leads to a negative magnetoresistance (MR)
 in presence of a weak uniform external magnetic field $B$.

Therefore, in the RMF problem one might expect that the semiclassical transport theory is applicable in a wider
range of length scales and 
temperatures than in the case of random time-reversal invariant potential scattering.
Yet, a semiclassical treatment of the RMF scattering beyond the relaxation time approximation may lead to a non-trivial MR.
 This classical contribution 
due to the bending of particle's trajectories dominates over the suppressed 
effects of weak localization, at least,  at not very low
temperatures.

The semiclassical approach to the RMF problem was undertaken in a number of publications. 
In \cite{MAW} the Subnikov-de Haas (SdH) oscillations of  $\rho_{xx}(B)$ in the strong  field limit $\Omega_c ={B\over m} >>{1/\tau_{tr}}$ were
studied by summing over classical cyclotron orbits. 
However, to find $\rho_{xx}(B)$ in the weak field limit before the onset of the SdH oscillations  one has to develop a semiclassical analysis in terms of particle-hole pairs rather than single fermions.

The authors of \cite{HS}   
used the linear Boltzmann equation where the RMF played a role of a random driving force instead
of including it into the collision integral. The condition which makes it possible is $k_F \xi>>1$, where $\xi$ is a correlation length (scale of a typical spatial variation)
of the RMF. However the analysis performed in \cite{HS} was restricted onto the case relevant for the experiments 
\cite{exp} where, first, strong potential scattering provides a large bare value of $\rho_{xx}$,
and, second, the spatial correlation of the RMF has a finite range $\sim \xi^{-1}$.

In the present letter we address the case  when $\rho_{xx}$
is solely due to the scattering by the RMF described by the (not necessarily short-range)
white-noise correlator 
$<b_q b_{-q}>=f(q)$ independent of the external field $B$.  We will not
restrict our consideration onto the lowest order of perturbation theory. Therefore we will be able  to discuss
 the case $f(q)\sim e^{-2q\xi}$ relevant for 
the problem of HFLL where the spatial correlation decays as $r^{-3}$ and the
lowest order result simply diverges.

We start with the Bolzmann equation for the distribution function $g(t, r, \phi,\epsilon )=g_0 (\epsilon)
 +\delta g(t, r, \phi, \epsilon)$ which reads as 
\begin{equation}
(i{\partial\over \partial t}+i{\vec v}_{\phi}{\vec \nabla}
+i(\Omega_c + {b({\vec r})\over m}){\partial\over \partial\phi})
\delta g(t, {\vec r}, \phi,\epsilon)=ie{\vec v}_{\phi}{\vec E}{\partial g_{0}\over \partial\epsilon},
\end{equation}
where ${\vec v}_{\phi}=v_F {\vec n}_{\phi}$ is a vector of Fermi velocity normal to the circular Fermi surface
parameterized by the angular variable $\phi$, ${\vec E}$ stands for an infinitesimal external
electric field, and $g_{0} = \theta(\varepsilon_F - \epsilon)$ is the unperturbed Fermi-Dirac distribution.
   By contrast to the case of ordinary potential scattering (which  can be also treated at 
$k_F \xi >>1$ as a random electric field) the Eq.(1) implies a trivial dependence of $\delta g$ on $\epsilon$,
namely,  
$\delta g(t, {\vec r}, \phi, \epsilon)=-\delta g(t, {\vec r}, \phi)\delta(\varepsilon_F - \epsilon)$.

We note, in passing, that one can use  the Eq.(1)  even in presence of long-range retarded gauge interactions between fermions
which spoil the 
Fermi liquid coherence. Even though there are no well-defined fermionic quasiparticles, 
the kinetic equation in terms of local displacements of the fluctuating Fermi surface can still be derived  \cite{L}, thereby making a close contact with the idea of bosonization of 2D non-Fermi liquids \cite{H}.

The perturbed distribution function can be found in terms of the (retarded) Green function of the Bolzmann
operator for a given RMF configuration    
\begin{equation}
\delta g(t, r, \phi,\epsilon )=ie \oint d\phi^{\prime}\int d{\vec r}^{\prime}
G(t; {\vec r}, {\vec r}^{\prime}; \phi, \phi^{\prime}){\vec v}_{\phi^{\prime}}{\vec E}\delta (\epsilon_F - \epsilon)
\end{equation}
and then averaged over all possible configurations.
As usual, the averaging over disorder restores the translational invariance in both $\vec r$- and $\phi$-spaces:
$<<G(t; {\vec r}, {\vec r}^{\prime}; \phi, \phi^{\prime})>>=
{\cal G}(t; {\vec r}-{\vec r}^{\prime}; \phi - \phi^{\prime})$.
Then the calculation of the frequency-dependent conductivity amounts to computing the $p$-wave harmonics of the $\vec q = 0$ component of the Fourier transform of
$$
{\cal G}(\omega; {\vec q}; \phi - \phi^{\prime})=\sum_{l} e^{il(\phi - \phi^{\prime}) + iR_c {\vec q}\times
({\vec n}_{\phi}-{\vec n}_{\phi^{\prime}})}{{\cal G}_{l}(\omega)\over 2\pi}
$$
where $R_c = {v_F\over \Omega_c}$ is the cyclotron radius of fermions with density $n_e$:
$$\sigma_{xx} (\omega)={ie^2 n_e\over 2m}({\cal G}_{1}(\omega)+{\cal G}_{-1}(\omega)), ~~~~~~~~~~~~
\sigma_{xy} (\omega)={e^2 n_e\over 2m}({\cal G}_{1}(\omega)-{\cal G}_{-1}(\omega))
$$
 In absence of the RMF the bare Green function is given by its
harmonics $G^{(0)}_{l}(\omega)={1\over {\omega - l\Omega_c + i\delta}}$.  

In our case of no potential disorder $(\delta=0^{+})$ the lowest order (bosonic) self-energy correction to ${\cal G}^{-1}_{l}(\omega)=
\omega - l\Omega_c + \Sigma_{l}(\omega)$ found in \cite{HS} makes it only possible  to study the $B$-dependent correction
to the bare $\rho_{xx}$ at high frequencies or magnetic fields. 
To proceed with a more complete account of the effects of the RMF 
we first solve the equation for the Green
function  
\begin{equation}
(i{\partial\over \partial t}+i{\vec v}_{\phi}{\vec \nabla}
+i(\Omega_c + {b({\vec r})\over m}){\partial\over \partial\phi}))
G(t-t^{\prime}; {\vec r}, {\vec r}^{\prime}; \phi, \phi^{\prime})
=\delta(t-t^{\prime})\delta({\vec r}- {\vec r}^{\prime})\delta(\phi- \phi^{\prime})
\end{equation}
for an arbitrary RMF configuration by Fourier transforming with respect to $t-t^\prime$ and ${\vec r}- {\vec r}^{\prime}$. The solution of the resulting equation 
\begin{equation}
(\omega -{\vec v}_{\phi}{\vec q} +i{\vec v}_{\phi}{\vec \nabla}
+i(\Omega_c + {b({\vec r})\over m}){\partial\over \partial\phi}))
G(\omega; {\vec q}, {\vec r} ; \phi, \phi^{\prime})=\delta(\phi- \phi^{\prime})
\end{equation}
can be searched in the form
\begin{equation}
G({\vec r}, {\vec q}; \phi, \phi^{\prime})=-i\int_{0}^{\infty}d\tau \oint {d\phi^{\prime\prime}}
e^{i\tau G_{0}^{-1}({\vec q}; \phi, \phi^{\prime\prime})}
e^{i\Psi(\tau; {\vec r}; \phi^{\prime\prime}, \phi^{\prime})}
\end{equation}
Here $\Psi$ can be viewed as a "eikonal" phase of a particle-hole pair propagating in the RMF.
Substituting (5) into (4) and integrating by parts we obtain the equation for $\Psi$:
\begin{equation}
(i{\partial\over \partial \tau}+i{\vec v}_{\phi}{\vec \nabla})\Psi=
- e^{-i\Psi}e^{-i\tau G_{0}^{-1}}
{b({\vec r})\over m}{\partial\over \partial\phi}e^{i\tau G_{0}^{-1}}e^{i\Psi}
\end{equation}
Provided the RMF correlation length is large enough $(k_F \xi >>1)$ one can linearize the Eq.(6) 
(we will comment on this point below) and then
end up with
an explicit solution 
\begin{equation}
\Psi(\tau; {\vec r}; \phi, \phi^{\prime})=
{1\over m}\int_{0}^{\tau}d\tau^{\prime}b({\vec r}-{\vec R}_{\tau^{\prime}}){\partial\over \partial\phi}
\delta(\phi- \phi^{\prime}),
\end{equation}
where  ${\vec R}_{\tau}=R_c e^{\Omega_c\tau{\partial\over \partial\phi}} {\vec n}_{\phi}$
is a classical trajectory corresponding to the Larmor precession along the cyclotron orbit.
Now taking the gaussian average of the exponent $e^{i\Psi}$ in (7) over different realizations of the RFM
we obtain the averaged Green function
 \begin{equation}
{\cal G}
({\vec r}, {\vec 0}; \phi - \phi^{\prime})=-i\int_{0}^{\infty}d\tau 
e^{i\tau (\omega +i\Omega_c {\partial\over \partial\phi})}
\exp({1\over 2}\int_{0}^{\tau}d\tau_{1}\int_{0}^{\tau}d\tau_{2}F(2R_c \sin{\Omega_c(\tau_{1}-\tau_{2})\over 2})
{\partial^2 \over \partial\phi^2})
 \delta(\phi- \phi^{\prime})
\end{equation}
where $F({ r})$ is a Fourier transform of the RMF correlator 
$f({q})$.
The use of the Eq. (8) yields the conductivity 
\begin{equation}
\sigma_{xx}(\omega)={e^2 n_e\over m}\int_{0}^{\infty}d\tau \cos{(\Omega_c+\omega)\tau}
\exp(-\int_{0}^{\tau}d\tau^{\prime}(\tau-\tau^{\prime})
F(2R_c \sin{\Omega_c\tau^{\prime}\over 2}))+(\omega\rightarrow -\omega)
\end{equation}
The formula for $\sigma_{xy}(\omega)$ differs from (9) by the additional factor $\tan(\Omega_c\pm\omega)\tau$ in the integrand.

The exponential factor in the integrand in (9) is controlled by the  
amplitude of the RMF correlator $F(r)$.
To check the validity of (9) in the perturbative regime we expand the exponent  up to the first order
in $F(r)$ and perform the $\tau$ integration first (it has to be done with an infinitesimal exponent
$e^{-\delta\tau}$ inserted into the integrand which specifies the retarded nature of the Green function (8)).
Then one can readily obtain
$$
\sigma_{xx}(\omega)={1/\tau_{tr}}(\Omega_c -\omega)^{-2}  + (\omega \rightarrow -\omega)
$$ 
where the RMF scattering rate is given by the expression
\begin{equation}
1/\tau_{tr}(B)=\int_{0}^{\infty}d\tau \cos{\Omega_c\tau}
F(2R_c \sin{\Omega_c\tau \over 2}),
\end{equation} 
which is nothing but the result of the first Born approximation (BA) \cite{HS}.
One can also obtain the formula (10) by
estimating the $\sim \tau$ term in the exponent in (9) in the $\tau \rightarrow \infty$ limit which would be equivalent to the relaxation time
approximation.

In what follows we will concentrate onto the case relevant for the problem of HFLL
corresponding to $f(q)= 4\pi^2\alpha n_e e^{-2q\xi}$ where we introduced the dimensionless coupling
$\alpha$ proportional to the density $n_i$ of ionized Coulomb impurities
 separated by the spacer of the width
$\xi$ from the 2D electron gas and the amount of gauge flux quanta $\Phi$ attached 
to every electron
($\alpha={1\over 2}\Phi^2 {n_i \over n_e}$ for a filling factor $\nu$ with an even denominator $\Phi$ \cite{HLR}).
In the picture of composite fermions (CF) \cite{HLR},\cite{KZ} each charged impurity also becomes a source
of a gauge magnetic flux. It is believed that scattering by randomly distributed fluxes provides the main
mechanism of the CF momentum relaxation while the potential scattering is negligible. 

At $B=0$ the Eq. (10) gives the CF elastic transport rate
 $1/\tau^{CF}_{tr}= {v_F \over 2\xi}\alpha$ which coincides with the BA result
found in \cite{HLR}.  Although this estimate certainly holds for $\alpha <<1$,
it is no longer valid at
 large $\alpha$ when the CF mean free path (MFP)  $l_{CF}=2\xi\alpha^{-1}$ gets shorter than $\xi$.
The case of HFLL $(\nu=1/2)$ appears to be marginal $(\alpha =2)$ and $l$ determined this way just equals $\xi$.
Since $\alpha$  is proportional to $\Phi^2$ the situation becomes even worse for compressible states at fractions with higher even denominators. 

This observation signals about a failure of the BA at large coupling $\alpha$ when the matrix element 
$M_{{\vec p},{\vec p}^{\prime}}\sim \sqrt\alpha {k_F\over m\xi}$
describing a single event of 
CF scattering by a typical magnetic impurity satisfies neither of the two conditions $M_{{\vec p},{\vec p}^{\prime}} <<{1\over m\xi^2}$ or ${(k_F \xi)\over m\xi^2}$ required
for the validity of the BA  \cite{LL}. This is the semiclassical regime of a small $(\Delta\phi\sim
{1\over {k_F \xi}}<<1)$
angle scattering,
which, however,  can not be treated in the lowest order of the perturbation theory.

Nevertheless, if the parameter $k_F\xi$ is large enough, so that
\begin{equation}
k_F \xi >> \sqrt\alpha,
\end{equation}
 one can resort on the so-called eikonal approach \cite{LL} which was essentially implemented by the above 
solution of the Boltzmann equation. 
In fact, it is the condition (11) which allows one to solve the equation (6) in the first order in 
$b({\vec r})$. 

The use of (9) then leads to
\begin{equation}
\sigma_{xx}= {e^2\over 2h}( k_F l)={e^2\over h}( k_F \xi)e^\alpha K_1 (\alpha),  
\end{equation}
where $K_1 (x)$ is the modified Bessel function of the second kind.
  At small $\alpha$ the Eq.(12) reproduces the above BA result while at 
strong coupling it predicts the MFP to be 
$l \approx \xi \sqrt{2\pi\over \alpha}$.

Thus the condition (11) actually provides that $k_F l >>1$ which is necessary to verify the very use
of the kinetic equation (1) and to obtain the metallic value of the 
conductivity.

In the intermediate coupling regime relevant for the case of HFLL the condition (11) is fairly well met ($\alpha=2$
and $k_F\xi\approx 15$) and
the Eq.(12) gives the CF conductivity $\sigma^{CF}_{xx}= {e^2\over h}( k_F \xi)e^2 K_{1}(2)$ which is 2.06 times greater
than the BA result.  
It is worthwhile mentioning that the experimentally measured resistivity at $\nu=1/2$ is about $3$ times smaller than the BA estimate \cite{HLR}. 

To add to this point we mention that a somewhat similar problem arises  in a semiclassical calculation of the amplitude of SdH 
oscilattions of the MR at large deviations from $\nu =1/2$ \cite{MAW}. The calculation carried out
in \cite{MAW} reproduces the experimentally observed scaling ${\Delta\rho^{osc}_{xx}\over \rho_{xx}}
\sim exp (-{const /B^{4}_{eff}})$ with the effective field $B_{eff}=B-4\pi n_e >> m/\tau^{CF}_{tr}$ but gives
a factor in the exponent which is about $2300$ times greater than the measured one. Presumably, such a discrepancy also stems from the inadequacy of the lowest order approximation
 (which systematically overestimates the strength
of the RMF scattering) and can be fixed by going beyond it.

One could wonder if the strong coupling behavior of the MFP as a function of the strength
of RMF fluctuations ($l\sim {\xi\over \sqrt{\alpha}}$ rather than $l\sim {\xi\over {\alpha}}$)
can be already seen in the self-consistent BA which in the case of a finite-range potential scattering is applicable  at $\xi<< B^{-1/2}$. 
Adapting this method to the RMF problem one can achieve a self-consistent improvement of the lowest order
result by inserting a factor $e^{-\tau/ \tau_{tr}}$ into the integrand in (10)
and solving the resulting nonlinear equation for $\tau_{tr}$.

For the case of HFLL this equation reads as
\begin{equation}
1/\tau_{tr}={2\xi\alpha\over v_F}
\int_{0}^{\infty}d\tau e^{-\tau/\tau_{tr}}(\tau^2 +({2\xi\over v_F})^2)^{-3/2}
\end{equation}
and gives the MFP $l(\alpha)$ in agreement with the above  
eikonal calculation. 
In terms of a conventional diagrammatics the equation (13) corresponds to the sum of uncrossed "rainbow" diagrams
for the bosonic self-energy $\Sigma_{1}$. By contrast, the eikonal result (8) includes contributions of some of the crossed
diagrams in all orders of the perturbation theory.

The MR in the RMF problem was identified in \cite{HS} as the $B$-dependence of $\tau_{tr}$. It is obvious, however, 
that in absence of other mechanisms of momentum relaxation one can not use the Eq.(10) to analyze the behavior of the MR at $\Omega_c < 1/\tau_{tr}$. 

Moreover, in the HFLL case the long-range character of $F(r)\sim r^{-3}$ 
leads to a logarithmic divergency of the second derivative of
 $\rho_{xx}(B)$ at $B\rightarrow 0$, which is, of course, an artifact of the above 
expansion. In the framework of the more accurate self-consistent BA one obtains that
in the weak field limit
$\tau_{tr}(B)$ is an increasing function of $B$ which 
implies a negative MR. Our self-consistent analysis shows that 
the decreasing behavior of $\tau_{tr}(B)$ (positive MR) in the case of $f(q)\sim e^{-{1\over 2}q^2 \xi^2}$ and small $\alpha$ reported in \cite{HS}, in fact,  holds only for large enough fields:
$\Omega_c > {1/\tau_{tr}}{1\over \log{1/\alpha^2}}$.

The more complete eikonal treatment of the RMF problem confirms
this prediction. Expanding (9) up to the second order in $\Omega_c$ we obtain the MR defined as 
${\Delta\rho_{xx}(B)\over \rho_{xx}}
=-{\Delta\sigma_{xx}(B)\over \sigma_{xx}}-({\sigma_{xy}(B)\over \sigma_{xx}})^2$ 
in the form
\begin{equation}
{\Delta\rho_{xx}(B)\over \rho_{xx}}
=
{\Omega_c^2\over 2<1>} (<\tau^2>-2{<\tau>^2 \over <1>} +{1\over 12}<\int_{0}^{\tau}d\tau^{\prime}
(3\tau^{\prime 2}\tau - 4\tau^{\prime 3})F(\tau^{\prime})>)
\end{equation}
where 
$$
<\tau^n>= \int_{0}^{\infty}d\tau \tau^n \exp(-\int_{0}^{\tau}d\tau^{\prime}(\tau-\tau^{\prime})F(\tau^{\prime}))
$$
Notice that in the relaxation time approximation $(<\tau^n>= \int_{0}^{\infty}d\tau \tau^n e^{-\tau/\tau_{tr}})$ the expression (14) contains only two terms proportional to $<\tau^2>$ and $<\tau>^2$ which exactly cancel out and result in zero MR.
Since the exponential factor $\int_{0}^{\tau}d\tau^{\prime}(\tau-\tau^{\prime})F(\tau^{\prime})$ appearing in all our calculations 
instead  of $\tau/\tau_{tr}$ behaves as $\sim\tau^2$ at
small $\tau$, the combined effect of these terms on the MR is negative and can dominate over the remaining (strictly positive)
contribution. 
 
The MR remains negative for all couplings in a wide class of realistic RMF correlation functions $f(q)$ including the HFLL case at $\alpha=2$ when the Eq.(14) yields 
$$
{\Delta\rho_{xx}(B)\over \rho_{xx}}=- 0.06 (\Omega_c\tau_{tr})^2
$$
while in the strong coupling limit of large $\alpha$ the MR becomes
${\Delta\rho_{xx}(B)\over \rho_{xx}}
=(\Omega_c\tau_{tr})^2 ({\pi\over 4}-1)$.

The possibility to apply our conclusions to real systems is complicated  by such factors as Fermi surface
and fermion dispersion anisotropy, periodic electron density modulations, or even a finite tilt of an external field which are
all known to yield positive contributions to the experimentally measured MR. 

The existing artificial realizations of the RMF \cite{exp} all have a property that the random field depends on the applied external one. A theoretical analysis shows that the  RMF correlation function $f(q)$ appropriate for the 
references (a) and (c) in \cite{exp} is proportional to $B$ while the one relevant for the reference (b) is $\sim B^2$. 
In either case this trivial $B$-dependence necessarily leads to a positive MR and does not allow one 
to explore a more sophisticated origin of a non-zero MR which is the only one present in the case of $f(q)$ independent of $B$.

The quadratic positive MR was also reported in a numerical simulation of the lattice version of the RMF modeled as an
assembly of uncorrelated random fluxes \cite{KWAZ}.
Again, the anisotropy of the lattice fermion dispersion and the obvious lack of the large
parameter $k_F \xi$ in the short-range case preclude us from making a direct comparison with our results.

In the case of HFLL one may also think of the observed positive MR as resulting from the
interference between the RMF scattering produced by charged impurities and strong gauge interactions
of CF. It was recently shown \cite{DVK} that this interference might be responsible for the strong non-universal $\log T$ correction to $\rho_{xx}(T)$ at $\nu =1/2$ \cite{log}. The account of the gauge interactions drastically changes the
renormalization properties of the RMF problem and can even reduce the localization length down to $L_{loc}\sim l\exp({\pi k_F l\over
{4\log k_F l}})$.

The zero field analogue of such an interference phenomenon (the Altshuler-Aronov correction) is known to lead in some cases to a
positive MR \cite{LR}. In our case of interest the effect of the external magnetic field could be even more pronounced.
The work intended to clarify this issue  is in progress.

To summarize, we develop a new  
method of solving the semiclassical Boltzmann equation for the RMF problem
beyond the relaxation time approximation provided $k_F l>>1$. The obtained solution  
is used to calculate the RMF transport time, mean free path, and zero-field conductivity which all appear to be 
greater than the corresponding results of the Born approximation. In particular, we propose a new estimate
of the semiclassical 
conductivity of composite fermions in the compressible state at $\nu=1/2$ which is about twice the value found in \cite{HLR} and 
agrees better with the experimental data. We also show that
in the ideal case of non-interacting fermions with uniform density and isotropic circular Fermi surface 
the magnetoresistance in the RMF is negative and quadratic in the weak external field.

The author is indebted to Per Hedegard for  
a valuable discussion of the results of this paper.

\pagebreak

\end{document}